\documentclass{article}
\usepackage{spconf,amsmath,graphicx}
\usepackage{graphicx}
\usepackage{siunitx}
\usepackage{color}
\usepackage[table]{xcolor} 
\usepackage{subcaption}
\usepackage{tabularx}
\usepackage{booktabs}
\usepackage{float}
\usepackage{multirow, array}

\usepackage{url}
\usepackage{xcolor}

\usepackage{hyperref}
\usepackage{amsmath}
\usepackage{comment}

\title{A two-step explainable approach for COVID-19 computer-aided diagnosis from chest x-ray images}

\name{
\begin{tabular}{c}
Carlo Alberto Barbano$^{\star}$ \qquad Enzo Tartaglione$^{\star}$ \qquad Claudio Berzovini$^{\dagger}$ \\ Marco Calandri$^{\ddagger}$ \qquad Marco Grangetto$^{\star}$\sthanks{This project has received funding from the European Union’s Horizon 2020 research and innovation programme under grant agreement No 825111, DeepHealth Project.}
\end{tabular}
}

\address{
    $^{\star}$ Computer Science Department, University of Turin, Italy \\
    $^{\dagger}$ Azienda Ospedaliera Città della Salute e della Scienza Presidio Molinette, Turin, Italy\\
    $^{\ddagger}$ Oncology Department, University of Turin, AOU San Luigi Gonzaga, Orbassano, Italy
}

\begin{document}
\onecolumn
\noindent © 20XX IEEE. Personal use of this material is permitted. Permission from IEEE must be obtained for all other uses, in any current or future media, including reprinting/republishing this material for advertising or promotional purposes, creating new collective works, for resale or redistribution to servers or lists, or reuse of any copyrighted component of this work in other works.
\twocolumn
\clearpage
\maketitle
\begin{abstract}
Early screening of patients is a critical issue in order to assess immediate and fast responses against the spread of COVID-19. The use of nasopharyngeal swabs has been considered the most viable approach; however, the result is not immediate or, in the case of fast exams, sufficiently accurate. Using Chest X-Ray (CXR) imaging for early screening potentially provides faster and more accurate response; however, diagnosing COVID from CXRs is hard and we should rely on deep learning support, whose decision process is, on the other hand, ``black-boxed'' and, for such reason, untrustworthy.\\
We propose an explainable two-step diagnostic approach, where we first detect known pathologies (anomalies) in the lungs, on top of which we diagnose the illness. Our approach achieves promising performance in COVID detection, compatible with expert human radiologists. All of our experiments have been carried out bearing in mind that, especially for clinical applications, explainability plays a major role for building trust in machine learning algorithms.
\end{abstract}


\section{Introduction}
\label{sec:introduction}
Early COVID diagnosis is a key element for proper treatment of the patients and prevention of the spread of the disease.
Given the high tropism of COVID-19 for respiratory airways and lung epythelium, identification of lung involvement in infected patients can be relevant for treatment and monitoring of the disease.
Virus testing is currently considered the only specific method of diagnosis. 
Nasopharingeal swabs are easily executable and affordable and current standard in diagnostic setting; their accuracy in literature is influenced by the severity of the disease and the time from symptoms onset and is reported up to 73.3\%~\cite{yang2020laboratory}. Current position papers from radiological societies (Fleischner Society, SIRM, RSNA)~\cite{ACR,SIRMpress,FLEISCHNER} do not recommend routine use of imaging for COVID-19 diagnosis; however, it has been widely demonstrated that, even at early stages of the disease, chest x-rays (CXR) can show pathological findings.\\

\begin{figure}
    \centering
    \includegraphics[width=1.0\columnwidth]{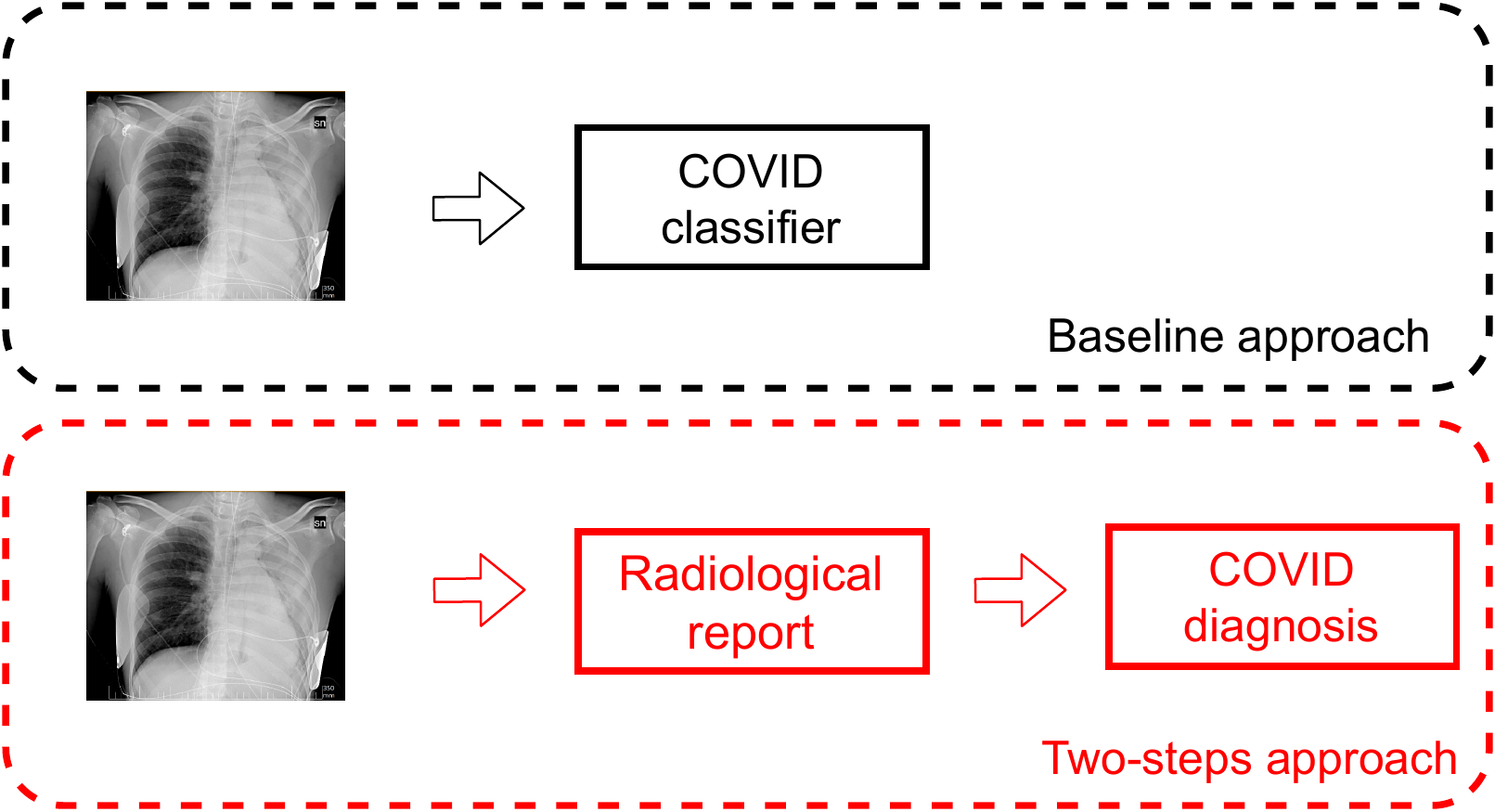}
    \caption{Comparison between standard approaches to COVID diagnosis and our two-step approach.} 
    \label{fig:our_approach}
\end{figure}

In the last year, many works attempted to tackle this problem, proposing deep learning-based strategies~\cite{tartaglione2020unveiling, sethy2020detection, apostolopoulos2020covid, narin2020automatic, wang2020covid}. 
All of the proposed approaches include some elements in common: i) the images collected during the pandemic need to be augmented with non-COVID cases from publicly available datasets; ii) some standard pre-processing is applied to the images, like lung segmentation using U-Net~\cite{ronneberger2015u} or similar models~\cite{tartaglione2020unveiling} or converting the pixels of the CXR scan in Hounsfield units; iii) the deep learning model is trained to the final diagnosis using state-of-the-art approaches for deep neural networks. 
Despite some very optimistic results, the proposed approaches exhibit significant limitations that deserve further analysis.  
For example, augmenting COVID datasets with negative cases from publicly-available datasets can inject a dangerous bias, where the trained model learn to discriminate different data sources rather than actual radiological features related to the disease~\cite{tartaglione2020unveiling}.  
These unwanted effects are difficult to spot when using a ``black box'' model like deep learning ones, without having control on the decision process.\\

\noindent In this work we propose an explainable approach, mimicking the radiologists' decision process. Towards this end, we break the COVID diagnosis problem into two sub-problems. First, we train a model to detect anomalies in the lungs. 
These anomalies are widely known and, following~\cite{hansell2008fleischner}, comprise 14 objective radiological observations which can be found in lungs. Then, on top of these, we train a decision tree model, where the COVID diagnosis is explicit (Fig.~\ref{fig:our_approach}). 
Mimicking the radiologist's decision is more robust to biases and aims at building trust for the physicians and patients towards the AI tool, which can be useful for fast COVID diagnosis. Thanks to the collaboration with the radiology units of Citt\`a della Salute e della Scienza di Torino (CDSS) and San Luigi Hospital (SLG) in Turin, we collected the COvid Radiographic images DAta-set for AI (CORDA), comprising both positive and negative COVID cases as well as a ground truth on the human radiological reporting, and it currently comprises almost 1000 CXRs. 

\section{Datasets}
\label{sec:datasets}

In this section we introduce the datasets that will be used for our proposed approach. 
\begin{figure}
    \centering
    \includegraphics[width=1.0\columnwidth]{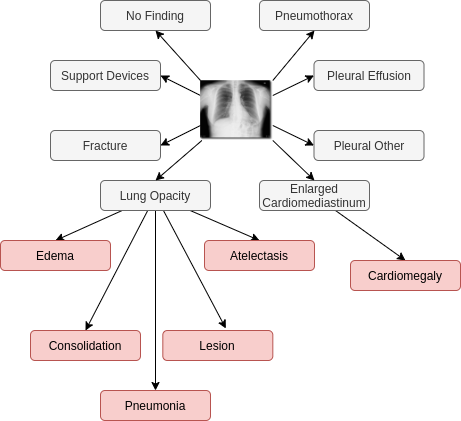}
    \caption{\emph{CheXpert}'s radiological findings.}
    \label{fig:chexpert-labels}
\end{figure}
For our purposes we first need to detect some objective radiological findings (we train a model on the \emph{CheXpert} dataset) and then, on top of those, we train a model to elaborate the COVID diagnosis (using the \emph{CORDA} dataset).\\

\noindent \textbf{CheXpert}: this is a large dataset comprising about 224k CXRs. This dataset consists of 14 different observations on the radiographic image: differently from many other datasets which are focused on disease classification based on clinical diagnosis, the main focus here is ``chest radiograph interpretation'', where anomalies are detected~\cite{irvin2019chexpert}. The learnable radiological findings are summarized in Fig.~\ref{fig:chexpert-labels}.\\

\noindent \textbf{CORDA}: this dataset was created for this study by retrospectively selecting chest x-rays performed at a dedicated Radiology Unit in CDSS and at SLG in all patients with fever or respiratory symptoms (cough, shortness of breath, dyspnea) that underwent nasopharingeal swab to rule out COVID-19 infection. Patients' average age is 61 years (range 17-97 years old). It contains a total of 898 CXRs and can be split by different collecting institution into two similarly sized subgroups: CORDA-CDSS~\cite{tartaglione2020unveiling}, which contains a total of 447 CXRs from 386 patients, with 150 images coming from COVID-negative patients and 297 from positive ones, and CORDA-SLG, which contains the remaining 451 CXRs, with 129 COVID-positive and 322 COVID-negative images. Including data from different  hospitals at test time is crucial to doublecheck the generalization capability of our model. 
The data collection is still in progress, with other 5 hospitals in Italy willing to contribute at time of writing. We plan to make CORDA available for research purposes according to EU regulations as soon as possible. 

\section{Radiological report}
\label{sec:radio-report}

In this section we are going to describe our proposed method to extract radiological findings from CXRs. For this task, we leverage the large scale dataset \emph{CheXpert}, which contains annotation for different kinds of common radiological findings that can be observed in CXR images (like opacity, pleural effusion, cardiomegaly, etc.). Given the high heterogeneity and the high cardinality of \emph{CheXpert}, its use is perfect for our purposes: in fact, once the model is trained on this dataset, there is no need to fine-tune it for the COVID diagnosis, since it will already extract objective radiological findings.\\
CheXpert provides 14 different types of observations for each image in the dataset. For each class, the labels have been generated from radiology reports associated with the studies with NLP techniques, conforming to the Fleischner Society’s recommended glossary~\cite{hansell2008fleischner}, and marked as: negative (N), positive (P), uncertain (U) or blank (N/A). 
Following the relationship among labels illustrated in Fig.~\ref{fig:chexpert-labels}, as proposed by \cite{irvin2019chexpert}, we can identify 8 top-level pathologies and 6 child ones.

\subsection{Dealing with uncertainty}
\begin{table*}
    \center
    \caption{Performance (AUC) for DenseNet-121 trained on CheXpert.}
    \label{table:results-dwu}
    \setlength{\tabcolsep}{7pt}
    \begin{tabular}{c c c c c c}
        \toprule
        \textbf{Method}                      &   \textbf{Atelectasis}     &   \textbf{Cardiomegaly}    &   \textbf{Consolidation}   &   \textbf{Edema}   &\textbf{Pleural Effusion}\\
        \hline
        Baseline~\cite{irvin2019chexpert}    & 0.79  &\textbf{0.81}  & 0.90  &0.91   &0.92 \\
        U-label use                 &\textbf{0.81}   &0.80           &\textbf{0.92}   &\textbf{0.94}   &\textbf{0.93}    \\
        \bottomrule
    \end{tabular}
\end{table*}
In order to extract the radiological findings from CXRs, a deep learning model is trained on the 14 observations. Towards this end, given the possibility of having multiple findings in the same CXR, the weighted binary cross entropy loss is used to train the model. Typically, weights are used to compensate class unbalancing, giving higher importance to less-represented classes. 
Within \emph{CheXpert}, however, we also need to tackle another issue: how to treat the samples with the U label. 
Towards this issue, multiple approaches have been suggested by~\cite{irvin2019chexpert}. The most popular is to ignore all the uncertain samples, excluding them from the training process and considering them as N/A.\\
We propose to include the U samples in the learning process, mapping them to maximum uncertainty (probability $0.5$ to be P or N). Then, we balance P and N outcomes for every radiological finding. Table~\ref{table:results-dwu} shows a performance comparison between the standard approach as proposed by~\cite{irvin2019chexpert} and our proposal (U-label use), for 5 salient radiological findings, using the same setting as in~\cite{irvin2019chexpert}. We observe an overall improvement in the performance, which is expected by the inclusion of the U-labeled examples. For all our experiments, we will use models trained using the U labeled samples.

\section{COVID diagnosis}
\label{sec:diagnosis}

The second step of the proposed approach is building the model which can actually provide a clinical diagnosis for COVID. We freeze the model obtained from Sec.~\ref{sec:radio-report} and use its output as image features to train a new binary classifier on the CORDA dataset. 
We test two different types of classifiers: a decision tree (Tree) and a neural network-based classifier (FC). \\
The decision tree is trained on the probabilities output of the radiological reports, using the state-of-the-art CART Algorithm implementation provided by the Python scikit-learn~\cite{scikit-learn} package.
Besides the fully explainable decision tree-based result, we also train a neural network classifier, comprising one hidden layer of size 512 and the output layer.
Despite working with the same features as the decision tree, such an approach loses in explainability, but potentially enhances the performance in terms of COVID diagnosis, as we will see in Sec.~\ref{sec:discussion}. 

\section{Results}
\label{sec:discussion}
\begin{table*}
    \center
    \caption{Results for COVID diagnosis.}
    \label{table:results-method1}
    \setlength{\tabcolsep}{5pt}
    \begin{tabular}{c c c c c c c c c c c c c}
        \toprule
        \textbf{Method}&\textbf{Backbone}&\textbf{Classifier}& \textbf{Pretrain dataset} & \textbf{Dataset} &
        \textbf{Sensitivity} & \textbf{Specificity} &
        \textbf{BA} & \textbf{AUC} \\
        \midrule
        \multirow{3}{*}{Baseline~\cite{tartaglione2020unveiling}}
                    &ResNet-18      &FC&none & CORDA-CDSS & \textbf{0.56} & 0.58 & 0.57 & 0.59 \\
                    &ResNet-18      &FC&RSNA & CORDA-CDSS & 0.54 & \textbf{0.80} & \textbf{0.67} & \textbf{0.72} \\
                    &ResNet-18      &FC&ChestXRay & CORDA-CDSS & 0.54 & 0.58 & 0.56 & 0.67 \\
        \midrule
                    
        \multirow{3}{*}{Two-step}   
         &ResNet-18      &FC&CheXpert& CORDA-CDSS & 0.69 & 0.73 & 0.71 & 0.76 \\
                    &DenseNet-121   &FC&CheXpert& CORDA-CDSS & 0.72 & \textbf{0.78} & \textbf{0.75} & \textbf{0.81} \\
                    &DenseNet-121   &Tree&CheXpert& CORDA-CDSS & \textbf{0.77} & 0.60 & 0.68 & 0.70 \\
        \midrule
        Two-step    &DenseNet-121   &FC&CheXpert& CORDA-SLG & 0.79 & 0.82 & 0.81 & 0.84 \\
        \bottomrule
    \end{tabular}
\end{table*}
In this section we compare the COVID diagnosis generalization capability through a direct deep learning-based approach (baseline) and our proposed two-step diagnosis, where first we detect the radiological findings, and then we discriminate patients affected by COVID using a decision tree-based diagnosis (Tree) or a deep learning-based classifier from the radiological findings (FC). The performance is tested on a subset of \emph{patients} not included in the training / validation set. The assessed metrics are: balanced accuracy (BA), sensitivity, specificity and area under the ROC curve (AUC). 
For all of the methods we adopt a 70\%-30\% train-test split. For the deep learning-based strategy, SGD is used with a learning rate $0.01$ and a weight decay of $10^{-5}$. 
\begin{figure*}
    \centering
    \includegraphics[width=0.95\textwidth, trim=37 37 37 37, clip]{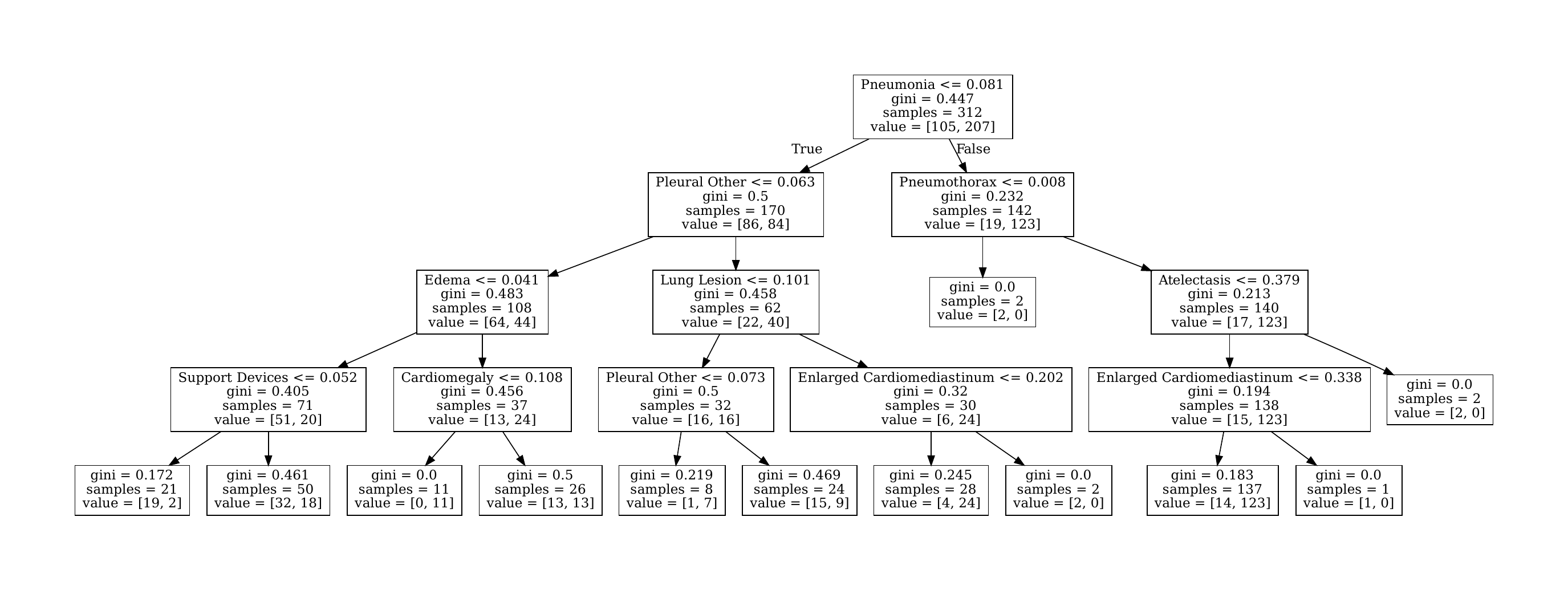}
    \caption{Decision Tree obtained for COVID-19 classification based on the probabilities for the 14 classes of findings.}
    \label{fig:densenet121-tree}
\end{figure*}

All of the experiments were run on NVIDIA Tesla T4 GPUs using PyTorch 1.4.
\noindent Table~\ref{table:results-method1} compares the standard deep learning-based approach~\cite{tartaglione2020unveiling} to our two-steps diagnosis.
Baseline results are obtained pre-training the model on some of the most used publicly-available datasets. We observe that the best achievable performance is very low, consisting in a BA of 0.67. A key takeaway is that trying to directly diagnose diseases such as COVID-19 from CXRs might be currently infeasible, probably given the small dataset sizes and strong selective bias in the datasets. \\
We can clearly see how the two-step method outperforms the direct diagnosis: using the same network architecture (ResNet-18 as backbone and a fully-connected classifier on top of it), we obtain a significant increase in all of the assessed metrics. Even better results are achieved by using a DenseNet-121 as backbone and the fully-connected classifier.

\noindent Fig.~\ref{fig:densenet121-tree} graphically shows the learned decision tree (whose performance is shown in Table~\ref{table:results-method1}): this provides a very clear interpretation for the decision process. 
From the clinical and radiological perspective, these data are consistent with the COVID-19 CXR semiotics that radiologists are used to deal with.  The edema feature, although unspecific, is strictly related to the interstitial involvement that is typical of COVID-19 infections and it has been largely reported in the recent literature~\cite{guan2020clinical}. Indeed, in recent COVID-19 radiological papers, interstitial involvement has been reported as ground glass opacity appearance~\cite{wong2020frequency}. However this definition is more pertinent to the CT imaging setting rather than CXR; the ``edema'' feature can be compatible, from the radiological perspective, to the interstitial opacity of COVID-19 patients. Furthermore, the not irrelevant role of cardiomegaly (or more in general enlarged cardiomediastinum) in the decision tree can be interesting from the clinical perspective. In fact, this can be read as an additional proof that established cardiovascular disease can be a relevant risk factor to develop COVID-19~\cite{guidance}.
Moreover, it may be consistent with the hypotheses of a larger role of the primary cardiovascular damage observed on on preliminary  data of autopsies of COVID-19 patients~\cite{wichmann2020autopsy}.

\begin{figure}
    \centering
    \begin{subfigure}{0.48\columnwidth}
        \includegraphics[width=\columnwidth]{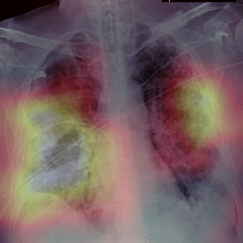}
    \end{subfigure}
    \begin{subfigure}{0.48\columnwidth}
        \includegraphics[width=\columnwidth]{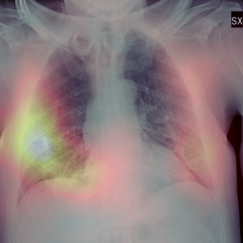}
    \end{subfigure}
    \\~\\
    \begin{subfigure}{0.48\columnwidth}
        \includegraphics[width=\columnwidth]{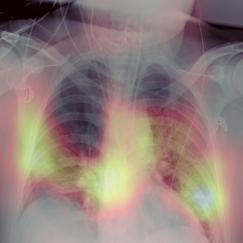}
    \end{subfigure}
    \begin{subfigure}{0.48\columnwidth}
        \includegraphics[width=\columnwidth]{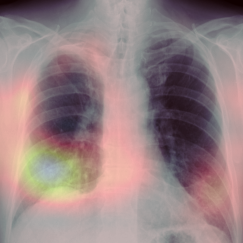}
    \end{subfigure}
    \caption{Grad-CAM on COVID-positive samples.}
    \label{fig:densenet121-gradcam}
\end{figure}
\noindent Focusing on the deep learning-based approach (FC) we observe a boost in the performance, achieving a BA of 0.75. However, this is the result of a trade-off between interpretability and discriminative power. Using Grad-CAM~\cite{selvaraju2017grad} we have hints on the area the model focused on to take the final diagnostic decision. From Fig.~\ref{fig:densenet121-gradcam} we observe that on COVID-positive images, the model seems to mostly focus on the expected lung areas.\\
Finally, to further test the reliability of our approach, we used our strategy also on CORDA-SLG (which are data coming from a different hospital structure), reaching comparable and encouraging results.
\section{Conclusions}
\label{sec:conclusions}
One of the latest challenges for both the clinical and the AI community has been applying deep learning in diagnosing COVID from CXRs. Recent works 
suggested the possibility of successfully tackling this problem, despite the currently small quantity of publicly available data.
In this work we propose a multi-step approach, close to the physicians' diagnostic process, in which the final diagnosis is based upon detected lung pathologies. 
We performed our experiments on CORDA, a COVID-19 CXR dataset comprising approximately 1000 images. 
All of our experiments have been carried out bearing in mind that, especially for clinical applications, explainability plays a major role for building trust in machine learning algorithms, although better interpretability can come at the cost of a lower prediction accuracy. 

\bibliographystyle{IEEEbib}
\bibliography{covidref}

\begin{thebibliography}{10}

\bibitem{yang2020laboratory}
Yang Yang, Minghui Yang, Chenguang Shen, Fuxiang Wang, Jing Yuan, Jinxiu Li,
  Mingxia Zhang, Zhaoqin Wang, Li~Xing, Jinli Wei, et~al.,
\newblock ``Laboratory diagnosis and monitoring the viral shedding of 2019-ncov
  infections,''
\newblock {\em medRxiv}, 2020.

\bibitem{ACR}
``{ACR} recommendations for the use of chest radiography and computed
  tomography ({CT}) for suspected {COVID-19} infection,''
  \href{https://www.acr.org/Advocacy-and-Economics/ACR-Position-Statements/Recommendations-for-Chest-Radiography-and-CT-for-Suspected-COVID19-Infection}{https://www.acr.org/}.

\bibitem{SIRMpress}
{Italian Radiology Society},
\newblock ``{Utilizzo della Diagnostica per Immagini nei pazienti Covid 19},''
  \href{https://www.sirm.org/wp-content/uploads/2020/03/DI-COVID-19-documento-intersocietario.pdf}{https://www.sirm.org/}.

\bibitem{FLEISCHNER}
Geoffrey~D. Rubin, Christopher~J. Ryerson, Linda~B. Haramati, Nicola
  Sverzellati, et~al.,
\newblock ``The role of chest imaging in patient management during the covid-19
  pandemic: A multinational consensus statement from the fleischner society,''
\newblock {\em RSNA Radiology}, 2020.

\bibitem{tartaglione2020unveiling}
Enzo Tartaglione, Carlo~Alberto Barbano, Claudio Berzovini, Marco Calandri, and
  Marco Grangetto,
\newblock ``Unveiling covid-19 from chest x-ray with deep learning: A hurdles
  race with small data,''
\newblock {\em International Journal of Environmental Research and Public
  Health}, vol. 17, no. 18, pp. 6933, Sep 2020.

\bibitem{sethy2020detection}
Prabira~Kumar Sethy and Santi~Kumari Behera,
\newblock ``Detection of coronavirus disease (covid-19) based on deep
  features,''
\newblock 2020.

\bibitem{apostolopoulos2020covid}
Ioannis~D Apostolopoulos and Tzani Bessiana,
\newblock ``Covid-19: Automatic detection from x-ray images utilizing transfer
  learning with convolutional neural networks,''
\newblock {\em arXiv preprint arXiv:2003.11617}, 2020.

\bibitem{narin2020automatic}
Ali Narin, Ceren Kaya, and Ziynet Pamuk,
\newblock ``Automatic detection of coronavirus disease (covid-19) using x-ray
  images and deep convolutional neural networks,''
\newblock {\em arXiv preprint arXiv:2003.10849}, 2020.

\bibitem{wang2020covid}
Linda Wang, Zhong~Qiu Lin, and Alexander Wong,
\newblock ``Covid-net: A tailored deep convolutional neural network design for
  detection of covid-19 cases from chest x-ray images,''
\newblock {\em Scientific Reports}, vol. 10, no. 1, pp. 1--12, 2020.

\bibitem{ronneberger2015u}
Olaf Ronneberger, Philipp Fischer, and Thomas Brox,
\newblock ``U-net: Convolutional networks for biomedical image segmentation,''
\newblock in {\em International Conference on Medical image computing and
  computer-assisted intervention}. Springer, 2015, pp. 234--241.

\bibitem{hansell2008fleischner}
David~M Hansell, Alexander~A Bankier, Heber MacMahon, Theresa~C McLoud,
  Nestor~L Muller, and Jacques Remy,
\newblock ``Fleischner society: glossary of terms for thoracic imaging,''
\newblock {\em Radiology}, vol. 246, no. 3, pp. 697--722, 2008.

\bibitem{irvin2019chexpert}
Jeremy Irvin, Pranav Rajpurkar, Michael Ko, Yifan Yu, Silviana Ciurea-Ilcus,
  Chris Chute, Henrik Marklund, Behzad Haghgoo, Robyn Ball, Katie Shpanskaya,
  et~al.,
\newblock ``Chexpert: A large chest radiograph dataset with uncertainty labels
  and expert comparison,''
\newblock in {\em Proceedings of the AAAI Conference on Artificial
  Intelligence}, 2019, vol.~33, pp. 590--597.

\bibitem{scikit-learn}
F.~Pedregosa, G.~Varoquaux, A.~Gramfort, V.~Michel, B.~Thirion, O.~Grisel,
  M.~Blondel, P.~Prettenhofer, R.~Weiss, V.~Dubourg, J.~Vanderplas, A.~Passos,
  D.~Cournapeau, M.~Brucher, M.~Perrot, and E.~Duchesnay,
\newblock ``Scikit-learn: Machine learning in {P}ython,''
\newblock {\em Journal of Machine Learning Research}, vol. 12, pp. 2825--2830,
  2011.

\bibitem{guan2020clinical}
Wei-jie Guan, Zheng-yi Ni, Yu~Hu, Wen-hua Liang, Chun-quan Ou, Jian-xing He,
  Lei Liu, Hong Shan, Chun-liang Lei, David~SC Hui, et~al.,
\newblock ``Clinical characteristics of coronavirus disease 2019 in china,''
\newblock {\em New England journal of medicine}, vol. 382, no. 18, pp.
  1708--1720, 2020.

\bibitem{wong2020frequency}
Ho~Yuen~Frank Wong, Hiu Yin~Sonia Lam, Ambrose Ho-Tung Fong, Siu~Ting Leung,
  Thomas Wing-Yan Chin, Christine Shing~Yen Lo, Macy Mei-Sze Lui, Jonan
  Chun~Yin Lee, Keith Wan-Hang Chiu, Tom Chung, et~al.,
\newblock ``Frequency and distribution of chest radiographic findings in
  covid-19 positive patients,''
\newblock {\em Radiology}, p. 201160, 2020.

\bibitem{guidance}
{\em ESC Guidance for the Diagnosis and Management of CV Disease during the
  COVID-19 Pandemic.}, 2020.

\bibitem{wichmann2020autopsy}
Dominic Wichmann, Jan-Peter Sperhake, Marc L{\"u}tgehetmann, Stefan Steurer,
  Carolin Edler, Axel Heinemann, Fabian Heinrich, Herbert Mushumba, Inga Kniep,
  Ann~Sophie Schr{\"o}der, et~al.,
\newblock ``Autopsy findings and venous thromboembolism in patients with
  covid-19: a prospective cohort study,''
\newblock {\em Annals of Internal Medicine}, 2020.

\bibitem{selvaraju2017grad}
Ramprasaath~R Selvaraju, Michael Cogswell, Abhishek Das, Ramakrishna Vedantam,
  Devi Parikh, and Dhruv Batra,
\newblock ``Grad-cam: Visual explanations from deep networks via gradient-based
  localization,''
\newblock in {\em Proceedings of the IEEE international conference on computer
  vision}, 2017, pp. 618--626.

\end{thebibliography}

\end{document}